\documentclass[twocolumn,amsmath,aps]{revtex4}

\usepackage{amssymb}
\usepackage{amsmath}
\usepackage{epsfig}
\usepackage{subfigure}
\usepackage{mathrsfs}
\usepackage{longtable}
\usepackage{float}
\usepackage{epstopdf}
\usepackage{hyperref}

\usepackage{graphicx}
\usepackage{float}
\usepackage{natbib}
\newcommand{\tab}{\hspace*{1em}}

\begin{document}

\title{A Lower Growth Rate from Recent Redshift Space Distortion Measurements than Expected from Planck}
\author{E. Macaulay}
\email{edward.macaulay@astro.ox.ac.uk}
\affiliation{Department of Physics \& Astronomy, University of Sussex, Falmer, BN1 9QH, UK}
\affiliation{Astrophysics, University of Oxford, Keble Road, Oxford, OX1 3RH, UK}
\author{I. K. Wehus}
\email{i.k.wehus@fys.uio.no}
\affiliation{Jet Propulsion Laboratory, California Institute of Technology, Pasadena,
CA 91109, USA}
\affiliation{Astrophysics, University of Oxford, DWB, Keble Road, Oxford, OX1 3RH, UK}
\author{H. K. Eriksen}
\email{h.k.k.eriksen@astro.uio.no}
\affiliation{Institute of Theoretical Astrophysics, University of Oslo, P.O.\ Box 1029 Blindern, N-0315 Oslo, Norway}

\begin{abstract}
We perform a meta-study of recently published Redshift Space Distortion (RSD) measurements of the cosmological growth rate, $f(z) \sigma_{8}(z)$.  We analyse the latest results from the 6dFGS, BOSS, LRG, WiggleZ and VIPERS galaxy redshift surveys, and compare the measurements to expectations from Planck.  In this Letter we point out that the RSD measurements are consistently lower than the values expected from Planck, and the relative scatter between the RSD measurements is lower than expected.  A full resolution of this issue may require a more robust treatment of non-linear effects in RSD models, although the trend for a low  $\sigma_{8}$ agrees with recent constraints on $\sigma_{8}$ and $\Omega_{m}$ from Sunyaev-Zeldovich cluster counts identified in Planck.
\end{abstract}

\maketitle

Understanding the accelerated expansion of the universe is currently one of the most important questions in cosmology.  Measurements of the distance-redshift relation with supernovae and Baryon Acoustic Oscillations (BAOs) are well described by General Relativity with a cosmological constant, and Cold Dark Matter -- the $\Lambda$CDM model.  The discovery of the accelerated expansion has motivated a vast number of theories of modified gravity - comprehensively reviewed by \cite{2012PhR...513....1C}.  Any theory of gravity must reproduce the background expansion observed with tests of the distance-redshift relation.  To test such theories, a number of galaxy surveys have measured the growth rate of cosmological density perturbations, where many modified gravity theories predict different growth rates to $\Lambda$CDM.  Specifically, the cosmological growth rate $f$ is defined as $f=d \ln G / d \ln  a$, where $a$ is the scale factor, and $G$ is the growth factor of the matter density contrast. 

Most recent growth rate measurements come from inferring peculiar velocities from Redshift Space Distortions (RSDs) in a galaxy redshift survey, as proposed by \cite{1987MNRAS.227....1K}.  One of the first RSD surveys to use this anisotropy to measure the growth rate was the 2dFGRS \citep{2001Natur.410..169P}.  The growth rate has since been measured with a range of other techniques and surveys, with the RSD technique in the VVDS survey \citep{2008Natur.451..541G}, QSO clustering and Ly$_{\alpha}$ clustering \citep{ross3812df,2008MNRAS.383..565D,2004MNRAS.354..684V}, and at $z \sim 0$ in peculiar velocity surveys, \citep{2011MNRAS.413.2906D,2012ApJ...751L..30H}. 

Since galaxies only form in the densest regions of the universe, a bias factor $b$ is used to relate perturbations in the matter density $\delta_{m}$ to perturbations in galaxy density $\delta_{g}$, so that $\delta_{g}=b\delta_{m}$.  Due to this bias, galaxies are only sensitive to the growth rate $f$ to within a factor of the density power spectrum normalisation.  Consequently, early growth rate measurements reported values of the parameter $\beta$, where $\beta=f/b$.  However, since the galaxy bias varies between populations of galaxies (with typical values between 1 and 3), values of $\beta$ from different surveys can be difficult to combine and compare to theories.  More recently, growth rate measurements have therefore been reported in the combination of $f(z) \sigma_{8}(z)$, \citep{2009MNRAS.393..297P} where $\sigma_{8}$ is the matter power spectrum normalisation on scales of 8 $h^{-1}$ Mpc.  It is only values of $f(z) \sigma_{8}(z)$ from RSD surveys that we consider here, as summarised in Table \ref{tab:GrowthRateData}, and not earlier values of $\beta$.

\begin{table*}[dtp]
\begin{center}
\begin{tabular}{llllllll}
Survey   & $z$ & $f(z) \sigma_{8}(z)$ & $r_{\text{max}}$ ($h^{-1}$Mpc) & Method & Model & Reference \\
\hline
6dFGS    & 0.067    &  $0.423\pm0.055$ & 30 & 0, 2, 4 & Scoccimarro \citep{2004PhRvD..70h3007S}  &  \cite{2012MNRAS.423.3430B} \\
LRG$_{200}$    & 0.25    &  $0.3512\pm0.0583$  &  200 & 0, 2, 4 & Kaiser \citep{1987MNRAS.227....1K}+damping &   \cite{2012MNRAS.420.2102S} \\
                      &  0.37    &  $0.4602\pm0.0378$     &   \\
LRG$_{60}$     & 0.25    &  $0.3665\pm0.0601$  &   60 & 0, 2, 4 & Kaiser \citep{1987MNRAS.227....1K}+damping   \\
                     &  0.37    &  $0.4031\pm0.0586$  &   \\
BOSS & 0.30 & $0.408\pm0.0552$ &  200 & 0, 2  & & \cite{2012MNRAS.424.2339T} \\
 $^{\rho=-0.19}$& 0.60 & $0.433\pm0.0662$ &  \\
 WiggleZ & 0.44 & $0.413\pm0.080$ &  $k_{\text{max}}$  & 0, 2  & Jennings \citep{2011MNRAS.410.2081J} & \cite{2012MNRAS.425..405B} \\
 $^{\rho=0.51}$& 0.60 & $0.390\pm0.063$   & 0.3 $h$Mpc$^{-1}$ &  \\
  $^{\rho=0.56}$& 0.73 & $0.437\pm0.072$  &  \\
VIPERS    & 0.8    &  $0.47\pm0.08$   &  30 & 0, 2 & Kaiser \citep{1987MNRAS.227....1K}+damping &  \cite{2013arXiv1303.2622D} \\
 \end{tabular}
\end{center}
\caption{Compilation of recent published values of $f(z) \sigma_{8}(z)$ (ordered by redshift).  Where the measurements within different redshift bins of the same survey are correlated, we indicate the correlation coefficient $\rho$ between the measurements (the 1$^{\text{st}}$ and 3$^{\text{rd}}$ redshift bins in the WiggleZ survey are uncorrelated). We also indicate the maximum scale used in the correlation function (the WiggleZ analysis uses the power spectrum), the indices of the Legendre moments used to fit for the anisotropic clustering \citep{1992ApJ...385L...5H}, and the model used to fit for the RSD.}

\label{tab:GrowthRateData}
\end{table*}

In \cite{2012MNRAS.424.2339T}, the growth rate from the BOSS survey was fitted at four correlated redshift values, although the publicly available covariance matrix is for three redshift measurements, to reduce correlations between the measurements.  We find that even with three redshift bins, the block-diagonal covariance matrix is too highly correlated, and thus we do not include the highly correlated intermediate redshift measurement.  We analyse the data with two different measurements from the LRG (Luminous Red Galaxy) survey (from the SDSS data release 7), with a maximum pair separation of 200 $h^{-1}$Mpc (LRG$_{200}$) and also with a maximum pair separation of 60 $h^{-1}$Mpc (LRG$_{60}$) -- we do not analyse the data with both  LRG$_{200}$ and LRG$_{60}$ simultaneously.   

\begin{figure}[t]
\begin{center}
\mbox{\epsfig{figure=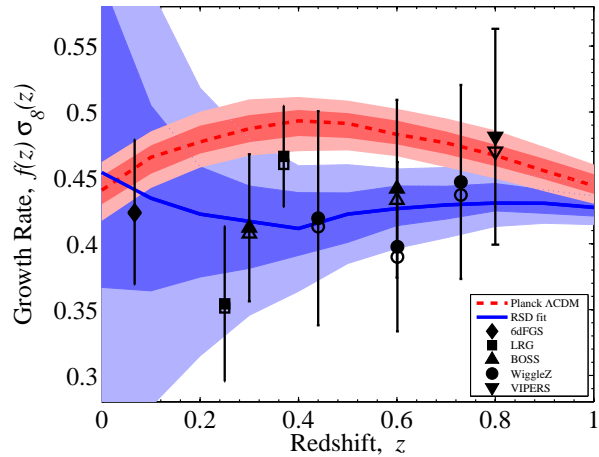,width=\linewidth,clip=.1}}
\end{center}
\caption{Comparing models to recent measurements of $f(z)\sigma_{8}(z)$.  We are plotting results for the LRG$_{200}$ data set.  The open markers are the original published values from the RSD measurements, and the filled markers are after accounting for the Alcock-Paczynski effect in going from WMAP to Planck cosmology.  The measurement error bars are at the 1 standard deviation uncertainty level.  The dashed red line illustrates the expected growth rate from $\Lambda$CDM with Planck parameters, with the 1 and 2 standard deviation uncertainty illustrated with the shaded bands.  The solid blue line and corresponding blue shaded regions illustrates the best fit to the RSD data with the gravitational slip model.  We note that almost all the measurements include our best fit model at the 1 standard deviation uncertainty level, which is reflected in the low $\chi^{2}$ in Table \ref{tab:GrowthRateResults}.  The one standard deviation range of the model (the darker blue band) is narrower than the typical one standard deviation uncertainty on any of the measurements because the fit has been calculated from the several independent measurements.}
\label{fig:f_sigma_eight}
\end{figure}

These galaxy surveys do not measure distances directly -- in order to infer the distance from the measured redshift, a cosmological model must be assumed.  As noted by \cite{1979Natur.281..358A}, if an incorrect cosmological model is assumed, an additional, artificial anisotropy can be imposed on the RSDs.  For the surveys we consider here, a $\Lambda$CDM cosmology based on Wilkinson Microwave Anisotropy Probe (WMAP) 7-year parameters \citep{2011ApJS..192...16L} was assumed.  Thus, in order to compare the measurements to predictions from Planck, we have to account for the additional anisotropy introduced by inferring distances from WMAP to Planck parameters.  To approximate this Alcock-Paczynski (AP) effect, we thus re-scale to growth rate measurements and uncertainties by the ratio of $H(z)D_{A}(z)$ in WMAP and Planck cosmologies, where $H(z)$ is the Hubble parameter, and $D_{A}(z)$ is the angular diameter distance.  In Figure \ref{fig:f_sigma_eight}, we plot the original published values of $f(z) \sigma_{8}(z)$ as open markers, and the re-scaled values as filled markers.

To account for the range of growth rate models allowed by Planck parameters, we use CAMB \citep{Lewis:1999bs} to generate growth rate models for each step in (a thinned version) of the Planck parameter chain.  To prepare the chain, we combine the eight \texttt{base\char`_planck\char`_lowl\char`_lowLike} chains from the Planck Legacy Archive \footnote{\url{http://www.sciops.esa.int/index.php?project=planck&page=Planck_Legacy_Archive}}, to create a chain 78,373 steps long, which -- for efficiency -- we thin by a factor of 10 to 7,838 steps.  These results are illustrated in Figure \ref{fig:f_sigma_eight}.  The lighter red band represents the region which includes 95\% of the growth rate models, and the darker red band illustrates the region which includes 68\%.  The dashed red line illustrates the best-fit.  To fit the RSD data, we thin the 7,838 step chain by an additional factor of 10, and for each of these steps perform a Markov chain Monte Carlo fit to the RSD data (calculating the AP-effect for the parameters at every step in the chain).  We then combine these chains to marginalise over the range of uncertainty allowed by Planck.

In order to fit the RSD data, we use parameters which -- as far as possible -- only affect the growth of perturbations, and not the well-constrained distance redshift relation or the Cosmic Microwave Background (CMB) anisotropies.  Following \cite{2012JCAP...06..032Z} we use a parameterised model for the gravitational slip, $\zeta$, given by
\begin{equation}
\Psi = ( 1 - \zeta ) \Phi
\end{equation}
where $\Psi$ and $\Phi$ are potentials which describe time-like and space-like metric perturbations in the Newtonian gauge, respectively.  For General Relativity, in the absence of anisotropic stress, these two potentials are equal, and so $\zeta=0$.  At the redshifts probed by RSDs, we expect the anisotropic stress to be negligible, so non-zero values of $\zeta$ may suggest physics beyond General Relativity.  For the particular model we consider here, we specify the value of $\zeta$ at redshift 0 and 1, we call these parameters $\zeta_{0}$ and $\zeta_{1}$.

The model additionally includes an equivalent parameterisation for an effective Newton's constant, although with only RSD data, the two sets of parameters are indistinguishable, and we consider only one set.  These parameters do not affect the background expansion, and only affects CMB anisotropies via the Integrated Sachs-Wolfe (ISW) effect.  In Table \ref{tab:GrowthRateResults} we illustrate the results of our fits for $\zeta_{0}$ and $\zeta_{1}$.  In both the LRG$_{60}$ and LRG$_{200}$ data sets, $\zeta_{0}$ is discrepant with the GR value of 0 at around the one standard deviation level, and $\zeta_{1}$ is discrepant at over the two standard deviation level.  The corresponding growth rate models are illustrated as before in Figure \ref{fig:f_sigma_eight} in blue, with a solid line.  

We note that the best fit to the RSD data would lead to a very high ISW signal in the CMB anisotropies.  On simultaneously fitting to low-$\ell$ CMB constraints (from WMAP) and RSDs -- and additionally fitting for an effective Newton's constant -- we find that the combined data is overwhelmingly dominated by the ISW constraint, only including the RSD data in the growth rate models at the 95\% limit.  Thus it does not currently appear possible to simultaneously fit RSDs and the ISW - the results we present here are for the fits to only the RSDs.

We consider the $\chi^2$ statistic for the fits, given by
\begin{equation}
\chi^2 = \left(x - \bar{x} \right) C^{-1} (x - \bar{x})
\end{equation}
where $x$ is a vector of observed values, $\bar{x}$ is a vector of corresponding values from a model for $x$, and $C$ is the covariance matrix for the data.  We note that for both data sets, the $\chi^2$ is substantially less than the 7 degrees of freedom in the fit.  We calculate the Probability To Exceed (PTE) this $\chi^2$, under the assumption that the uncertainties are indeed correctly estimated.  The very low PTE values suggest that either the uncertainties have been over estimated, or genuine scatter in the measurements is being systematically suppressed.   While only additional observations will determine whether this trend is truly statistically significant, the results already in hand appear to suggest that either the quoted uncertainties have been overestimated, or the analysis is suppressing genuine scatter in the measurements. 

\begin{table}[dtp]
\begin{center}
\begin{tabular}{c c c c c}
$\zeta_{0}$  \tab& $\zeta_{1}$  \tab& $\rho$  \tab& $\chi^2$  \tab& 1-PTE  \\
\hline
-2.94$\pm$1.94	  \tab&  0.32$\pm$0.13  \tab& -0.72 \tab& 1.34 \tab& 0.99 \\
-2.07$\pm$1.88  \tab&  0.28$\pm$0.10  \tab& -0.70 \tab& 3.31 \tab& 0.86 \\

 \end{tabular}
\end{center}
\caption{Results from fits to the RSD data.  The first line of results is for the LRG$_{60}$ data set, and the second line is for LRG$_{200}$.  For each set, we present the best-fit values of the gravitational slip at redshift 0 and 1 ($\zeta_{0}$ \& $\zeta_{1}$).  The uncertainties are at the one-standard deviation level.  The fiducial value of both parameters in General Relativity is 0.  We also indicate the correlation coefficient $\rho$ of the distribution of the fit to these two parameters, the minimum $\chi^2$ of the fit and corresponding Probability To Exceed (PTE).}
\label{tab:GrowthRateResults}
\end{table}

We note that the PTE decreases with the LRG$_{200}$ data set, since the LRG$_{200}$ measurements have a larger scatter than the LRG$_{60}$ measurements.  This is likely due to the fact that most of the coherent clustering signal is due to correlations on scales less than 100 $h^{-1}$Mpc, so the additional correlations are effectively adding noise to the signal.  

In most recent results, the uncertainties have been estimated from several hundred simulated realisations of the survey, from which the uncertainty (and the covariance between measurements, in the case of several redshift bins) can be deduced from the scatter in the realisations.  Although it may appear that the uncertainties on the measurements have been overestimated, good agreement between the quoted values and Fisher forecasts \citep[e.g.,][]{2009MNRAS.397.1348W} of the minimum intrinsic statistical uncertainties suggests that this is not the case, although \cite{2012MNRAS.426.2719R} note that the uncertainties in the BOSS growth rate measurements are around 40\% larger than the Fisher matrix predictions.  

Perhaps the stage of an RSD analysis most likely to introduce a systematic shift, and artificially reduce the scatter, may be in fitting a model to the two-dimensional two-point correlation function (or power spectrum).  \cite{2012arXiv1210.2130P} analysed simulated catalogues for the WiggleZ survey with a range of models for the RSD effect, and found that measurements of $\Omega_{m}$ (which is directly sensitive to the growth rate) were highly dependent on the model used.  In particular, the model of a HALOFIT \citep{2003MNRAS.341.1311S} $P(k)$ with a linear model for the redshift space distortion recovered a lower $\Omega_{m}$ compared to the fiducial value on which the simulation was based.  

 The preference for a lower growth rate or $\sigma_{8}$ appears to agree with recent results from \cite{2013arXiv1303.5080P}, studying Sunyaev-Zeldovich (SZ) cluster counts, who find $\sigma_{8}=0.77\pm0.02$ and $\Omega_{m}=0.29\pm0.02$.  Collectively, these results may be suggesting that $\Lambda$CDM does not fully model simultaneously the Cosmic Microwave Background and the Universe at $z<1$.  However, future work will require detailed work with simulated catalogues for a range of cosmological models \citep[e.g.,][]{2012MNRAS.tmpL.533J,2012MNRAS.425.2128J} and an improved understanding of the relationship between the observed galaxies, the peculiar velocity field, and the underlying dark matter \citep[e.g.,][]{2010Natur.464..256R,2012PhRvD..86h3006S}, before we can more robustly use RSD measurements to study departures from $\Lambda$CDM.

We thank the two anonymous referees for useful comments which have substantially improved this letter, Jonathan Patterson for help with parallelisation of the analysis code, Tessa Baker for help with the model parametrisation, Erminia Calabrese for help with the Planck parameters for the growth rate, Lado Samushia for help with the growth rate models, Rita Tojeiro for help with the BOSS data, Chris Blake for useful suggestions and help with the WiggleZ data and Pedro Ferreira for comments and discussions.  This project was supported by an ERC Starting Grant StG2010-257080 and a Leverhulme visiting professorship for HKE.  EM acknowledges support from the BIPAC and STFC.   IKW acknowledges support from ERC grant 259505.  Part of the research was carried out at the Jet Propulsion Laboratory, California Institute of Technology, under a contract with NASA.

\bibliographystyle{hapj}
\bibliography{growthrate_refs}

\begin{thebibliography}{31}
\expandafter\ifx\csname natexlab\endcsname\relax\def\natexlab#1{#1}\fi

\bibitem[{{Alcock} \& {Paczynski}(1979)}]{1979Natur.281..358A}
{Alcock}, C., \& {Paczynski}, B. 1979, \nat, 281, 358

\bibitem[{{Beutler} {et~al.}(2012){Beutler}, {Blake}, {Colless}, {Jones},
  {Staveley-Smith}, {Poole}, {Campbell}, {Parker}, {Saunders}, \&
  {Watson}}]{2012MNRAS.423.3430B}
{Beutler}, F. {et~al.} 2012, \mnras, 423, 3430, 1204.4725

\bibitem[{{Blake} {et~al.}(2012){Blake}, {Brough}, {Colless}, {Contreras},
  {Couch}, {Croom}, {Croton}, {Davis}, {Drinkwater}, {Forster}, {Gilbank},
  {Gladders}, {Glazebrook}, {Jelliffe}, {Jurek}, {Li}, {Madore}, {Martin},
  {Pimbblet}, {Poole}, {Pracy}, {Sharp}, {Wisnioski}, {Woods}, {Wyder}, \&
  {Yee}}]{2012MNRAS.425..405B}
{Blake}, C. {et~al.} 2012, \mnras, 425, 405, 1204.3674

\bibitem[{{Clifton} {et~al.}(2012){Clifton}, {Ferreira}, {Padilla}, \&
  {Skordis}}]{2012PhR...513....1C}
{Clifton}, T., {Ferreira}, P.~G., {Padilla}, A., \& {Skordis}, C. 2012,
  \physrep, 513, 1, 1106.2476

\bibitem[{{da {\^A}ngela} {et~al.}(2008){da {\^A}ngela}, {Shanks}, {Croom},
  {Weilbacher}, {Brunner}, {Couch}, {Miller}, {Myers}, {Nichol}, {Pimbblet},
  {de Propris}, {Richards}, {Ross}, {Schneider}, \&
  {Wake}}]{2008MNRAS.383..565D}
{da {\^A}ngela}, J. {et~al.} 2008, \mnras, 383, 565, arXiv:astro-ph/0612401

\bibitem[{{Davis} {et~al.}(2011){Davis}, {Nusser}, {Masters}, {Springob},
  {Huchra}, \& {Lemson}}]{2011MNRAS.413.2906D}
{Davis}, M., {Nusser}, A., {Masters}, K.~L., {Springob}, C., {Huchra}, J.~P.,
  \& {Lemson}, G. 2011, \mnras, 413, 2906, 1011.3114

\bibitem[{{de la Torre} {et~al.}(2013){de la Torre}, {Guzzo}, {Peacock},
  {Branchini}, {Iovino}, {Granett}, {Abbas}, {Adami}, {Arnouts}, {Bel},
  {Bolzonella}, {Bottini}, {Cappi}, {Coupon}, {Cucciati}, {Davidzon}, {De
  Lucia}, {Fritz}, {Franzetti}, {Fumana}, {Garilli}, {Ilbert}, {Krywult}, {Le
  Brun}, {Le Fevre}, {Maccagni}, {Malek}, {Marulli}, {McCracken}, {Moscardini},
  {Paioro}, {Percival}, {Polletta}, {Pollo}, {Schlagenhaufer}, {Scodeggio},
  {Tasca}, {Tojeiro}, {Vergani}, {Zanichelli}, {Burden}, {Di Porto},
  {Marchetti}, {Marinoni}, {Mellier}, {Monaco}, {Nichol}, {Phleps}, {Wolk}, \&
  {Zamorani}}]{2013arXiv1303.2622D}
{de la Torre}, S. {et~al.} 2013, ArXiv e-prints, 1303.2622

\bibitem[{{Guzzo} {et~al.}(2008){Guzzo}, {Pierleoni}, {Meneux}, {Branchini},
  {Le F{\`e}vre}, {Marinoni}, {Garilli}, {Blaizot}, {De Lucia}, {Pollo},
  {McCracken}, {Bottini}, {Le Brun}, {Maccagni}, {Picat}, {Scaramella},
  {Scodeggio}, {Tresse}, {Vettolani}, {Zanichelli}, {Adami}, {Arnouts},
  {Bardelli}, {Bolzonella}, {Bongiorno}, {Cappi}, {Charlot}, {Ciliegi},
  {Contini}, {Cucciati}, {de la Torre}, {Dolag}, {Foucaud}, {Franzetti},
  {Gavignaud}, {Ilbert}, {Iovino}, {Lamareille}, {Marano}, {Mazure}, {Memeo},
  {Merighi}, {Moscardini}, {Paltani}, {Pell{\`o}}, {Perez-Montero}, {Pozzetti},
  {Radovich}, {Vergani}, {Zamorani}, \& {Zucca}}]{2008Natur.451..541G}
{Guzzo}, L. {et~al.} 2008, \nat, 451, 541, 0802.1944

\bibitem[{{Hamilton}(1992)}]{1992ApJ...385L...5H}
{Hamilton}, A.~J.~S. 1992, \apjl, 385, L5

\bibitem[{{Hudson} \& {Turnbull}(2012)}]{2012ApJ...751L..30H}
{Hudson}, M.~J., \& {Turnbull}, S.~J. 2012, \apjl, 751, L30, 1203.4814

\bibitem[{{Jennings}(2012)}]{2012MNRAS.tmpL.533J}
{Jennings}, E. 2012, \mnras, L533, 1207.1439

\bibitem[{{Jennings} {et~al.}(2012){Jennings}, {Baugh}, {Li}, {Zhao}, \&
  {Koyama}}]{2012MNRAS.425.2128J}
{Jennings}, E., {Baugh}, C.~M., {Li}, B., {Zhao}, G.-B., \& {Koyama}, K. 2012,
  \mnras, 425, 2128, 1205.2698

\bibitem[{{Jennings} {et~al.}(2011){Jennings}, {Baugh}, \&
  {Pascoli}}]{2011MNRAS.410.2081J}
{Jennings}, E., {Baugh}, C.~M., \& {Pascoli}, S. 2011, \mnras, 410, 2081,
  1003.4282

\bibitem[{{Kaiser}(1987)}]{1987MNRAS.227....1K}
{Kaiser}, N. 1987, \mnras, 227, 1

\bibitem[{{Larson} {et~al.}(2011){Larson}, {Dunkley}, {Hinshaw}, {Komatsu},
  {Nolta}, {Bennett}, {Gold}, {Halpern}, {Hill}, {Jarosik}, {Kogut}, {Limon},
  {Meyer}, {Odegard}, {Page}, {Smith}, {Spergel}, {Tucker}, {Weiland},
  {Wollack}, \& {Wright}}]{2011ApJS..192...16L}
{Larson}, D. {et~al.} 2011, \apjs, 192, 16, 1001.4635

\bibitem[{Lewis {et~al.}(2000)Lewis, Challinor, \& Lasenby}]{Lewis:1999bs}
Lewis, A., Challinor, A., \& Lasenby, A. 2000, Astrophys. J., 538, 473,
  astro-ph/9911177

\bibitem[{{Parkinson} {et~al.}(2012){Parkinson}, {Riemer-S{\o}rensen}, {Blake},
  {Poole}, {Davis}, {Brough}, {Colless}, {Contreras}, {Couch}, {Croom},
  {Croton}, {Drinkwater}, {Forster}, {Gilbank}, {Gladders}, {Glazebrook},
  {Jelliffe}, {Jurek}, {Li}, {Madore}, {Martin}, {Pimbblet}, {Pracy}, {Sharp},
  {Wisnioski}, {Woods}, {Wyder}, \& {Yee}}]{2012arXiv1210.2130P}
{Parkinson}, D. {et~al.} 2012, ArXiv e-prints, 1210.2130

\bibitem[{{Peacock} {et~al.}(2001){Peacock}, {Cole}, {Norberg}, {Baugh},
  {Bland-Hawthorn}, {Bridges}, {Cannon}, {Colless}, {Collins}, {Couch},
  {Dalton}, {Deeley}, {De Propris}, {Driver}, {Efstathiou}, {Ellis}, {Frenk},
  {Glazebrook}, {Jackson}, {Lahav}, {Lewis}, {Lumsden}, {Maddox}, {Percival},
  {Peterson}, {Price}, {Sutherland}, \& {Taylor}}]{2001Natur.410..169P}
{Peacock}, J.~A. {et~al.} 2001, \nature, 410, 169, arXiv:astro-ph/0103143

\bibitem[{{Percival} \& {White}(2009)}]{2009MNRAS.393..297P}
{Percival}, W.~J., \& {White}, M. 2009, \mnras, 393, 297, 0808.0003

\bibitem[{{Planck Collaboration} {et~al.}(2013){Planck Collaboration}, {Ade},
  {Aghanim}, {Armitage-Caplan}, {Arnaud}, {Ashdown}, {Atrio-Barandela},
  {Aumont}, {Baccigalupi}, {Banday}, \& et~al.}]{2013arXiv1303.5080P}
{Planck Collaboration} {et~al.} 2013, ArXiv e-prints, 1303.5080

\bibitem[{{Reid} {et~al.}(2012){Reid}, {Samushia}, {White}, {Percival},
  {Manera}, {Padmanabhan}, {Ross}, {S{\'a}nchez}, {Bailey}, {Bizyaev},
  {Bolton}, {Brewington}, {Brinkmann}, {Brownstein}, {Cuesta}, {Eisenstein},
  {Gunn}, {Honscheid}, {Malanushenko}, {Malanushenko}, {Maraston}, {McBride},
  {Muna}, {Nichol}, {Oravetz}, {Pan}, {de Putter}, {Roe}, {Ross}, {Schlegel},
  {Schneider}, {Seo}, {Shelden}, {Sheldon}, {Simmons}, {Skibba}, {Snedden},
  {Swanson}, {Thomas}, {Tinker}, {Tojeiro}, {Verde}, {Wake}, {Weaver},
  {Weinberg}, {Zehavi}, \& {Zhao}}]{2012MNRAS.426.2719R}
{Reid}, B.~A. {et~al.} 2012, \mnras, 426, 2719, 1203.6641

\bibitem[{{Reyes} {et~al.}(2010){Reyes}, {Mandelbaum}, {Seljak}, {Baldauf},
  {Gunn}, {Lombriser}, \& {Smith}}]{2010Natur.464..256R}
{Reyes}, R., {Mandelbaum}, R., {Seljak}, U., {Baldauf}, T., {Gunn}, J.~E.,
  {Lombriser}, L., \& {Smith}, R.~E. 2010, \nat, 464, 256, 1003.2185

\bibitem[{Ross {et~al.}(2007)Ross, da~Angela, Shanks, Wake, Cannon, Edge,
  Nichol, Outram, Colless, Couch, {et~al.}}]{ross3812df}
Ross, N. {et~al.} 2007, \mnras, 381, 573

\bibitem[{{Samushia} {et~al.}(2012){Samushia}, {Percival}, \&
  {Raccanelli}}]{2012MNRAS.420.2102S}
{Samushia}, L., {Percival}, W.~J., \& {Raccanelli}, A. 2012, \mnras, 420, 2102,
  1102.1014

\bibitem[{{Scoccimarro}(2004)}]{2004PhRvD..70h3007S}
{Scoccimarro}, R. 2004, \prd, 70, 083007, arXiv:astro-ph/0407214

\bibitem[{{Sherwin} {et~al.}(2012){Sherwin}, {Das}, {Hajian}, {Addison},
  {Bond}, {Crichton}, {Devlin}, {Dunkley}, {Gralla}, {Halpern}, {Hill},
  {Hincks}, {Hughes}, {Huffenberger}, {Hlozek}, {Kosowsky}, {Louis},
  {Marriage}, {Marsden}, {Menanteau}, {Moodley}, {Niemack}, {Page}, {Reese},
  {Sehgal}, {Sievers}, {Sif{\'o}n}, {Spergel}, {Staggs}, {Switzer}, \&
  {Wollack}}]{2012PhRvD..86h3006S}
{Sherwin}, B.~D. {et~al.} 2012, \prd, 86, 083006, 1207.4543

\bibitem[{{Smith} {et~al.}(2003){Smith}, {Peacock}, {Jenkins}, {White},
  {Frenk}, {Pearce}, {Thomas}, {Efstathiou}, \&
  {Couchman}}]{2003MNRAS.341.1311S}
{Smith}, R.~E. {et~al.} 2003, \mnras, 341, 1311, arXiv:astro-ph/0207664

\bibitem[{{Tojeiro} {et~al.}(2012){Tojeiro}, {Percival}, {Brinkmann},
  {Brownstein}, {Eisenstein}, {Manera}, {Maraston}, {McBride}, {Muna}, {Reid},
  {Ross}, {Ross}, {Samushia}, {Padmanabhan}, {Schneider}, {Skibba},
  {S{\'a}nchez}, {Swanson}, {Thomas}, {Tinker}, {Verde}, {Wake}, {Weaver}, \&
  {Zhao}}]{2012MNRAS.424.2339T}
{Tojeiro}, R. {et~al.} 2012, \mnras, 424, 2339, 1203.6565

\bibitem[{{Viel} {et~al.}(2004){Viel}, {Haehnelt}, \&
  {Springel}}]{2004MNRAS.354..684V}
{Viel}, M., {Haehnelt}, M.~G., \& {Springel}, V. 2004, \mnras, 354, 684,
  arXiv:astro-ph/0404600

\bibitem[{{White} {et~al.}(2009){White}, {Song}, \&
  {Percival}}]{2009MNRAS.397.1348W}
{White}, M., {Song}, Y.-S., \& {Percival}, W.~J. 2009, \mnras, 397, 1348,
  0810.1518

\bibitem[{{Zuntz} {et~al.}(2012){Zuntz}, {Baker}, {Ferreira}, \&
  {Skordis}}]{2012JCAP...06..032Z}
{Zuntz}, J., {Baker}, T., {Ferreira}, P.~G., \& {Skordis}, C. 2012, \jcap, 6,
  32, 1110.3830

\end{thebibliography}
\end{document}